\documentclass[twocolumn,floatfix,amsmath,amssymb,aps,prl]{revtex4-1}

\usepackage[utf8]{inputenc}
\usepackage{enumitem,dsfont,bm,bbold}
\usepackage{hyperref}
\usepackage{array,dcolumn,multirow,braket}
\usepackage{graphicx,xcolor}
\graphicspath{{./}{./figures/}}

\def\O{ {\mathrm{O}} }

\def\vk{ {\boldsymbol{k}} }

\newcommand{\beq}{\begin{equation}}
\newcommand{\eeq}{\end{equation}}

\newcommand{\abs}[1]{\left| #1 \right|}

\newcommand{\nullv}{\mathbf{0}}

\newcommand{\diag}[1]{\text{diag}\left\{ #1 \right\}}

\renewcommand{\O}{\mathrm{O}}

\newcommand{\ve}{\varepsilon}

\newcommand{\hlt}{\mathcal{H}}

\newcommand{\pd}{{\phantom{\dagger}}}
\newcommand{\tht}{\tilde{\theta}}

\begin{document}

\title{Magnetic Breakdown and Chiral Magnetic Effect at \\ Weyl-Semimetal Tunnel Junctions}

\author{Adam Yanis Chaou}
\author{Vatsal Dwivedi}
\author{Maxim Breitkreiz}
\email{breitkr@physik.fu-berlin.de}

\affiliation{Dahlem Center for Complex Quantum Systems and Fachbereich Physik, Freie Universit\" at Berlin, 14195 Berlin, Germany}


\begin{abstract}
We investigate magnetotransport across an interface between two Weyl semimetals whose Weyl nodes project onto different interface momenta. Such an interface generically hosts Fermi arcs that connect Weyl nodes of identical chirality in different Weyl semimetals (homochiral connectivity) --- in contrast to surface Fermi arcs that connect opposite-chirality Weyl nodes within the same Weyl semimetal (heterochiral connectivity). We show that electron transport along the arcs with homochiral connectivity, in the presence of a longitudinal magnetic field, leads to a universal longitudinal magnetoconductance of $e^2/h$ per magnetic flux quantum. Furthermore, a weak tunnel coupling can result in a close encounter of two homochiral-connectivity Fermi arcs, enabling magnetic breakdown. Above the breakdown field the interface Fermi arc connectivity is effectively heterochiral, leading to a saturation of the conductance. 
\end{abstract}

\maketitle

\textit{Introduction} --- 
Weyl semimetals (WSMs) are a class of three dimensional semimetals characterized by pairs of opposite-chirality Weyl fermions instantiated as topologically protected gapless points in the bulk Brillouin zone (BZ) \cite{Wan2011,Burkov2011,Xu2011,Xu2015, Xu2015b, Lv2015,Armitage2017, Yan2017, Burkov2017,Hasan2021, Bernevig2022a}. Individual Weyl fermions exhibit the chiral anomaly \cite{Adler1969,Bell1969}  --- a violation of particle-number conservation in presence of parallel electric and magnetic fields. The chiral anomaly manifests as a spectral flow on chiral zeroth Landau levels, which disperse either parallel or anti-parallel to an applied magnetic field depending on chirality of the Weyl fermion \cite{Nielsen1983}. In a finite system, the reconnection of this spectral flow necessitates the existence of gapless Fermi-arc surface states, which connect Weyl nodes of opposite chirality. 

One of the most striking transport phenomena associated with the chiral anomaly is the \textit{chiral magnetic effect} (CME), which leads to a positive longitudinal magnetoconductance \cite{Son2013a, Burkov2017, Xiong2015a}. In the ballistic (also called ultra-quantum) limit, where transport is governed solely by the lowest (chiral) Landau level, the longitudinal conductance is predicted to show a universal linear dependence pm the magnetic field  \cite{Altland2016}. Experimental evidence of the chiral anomaly by way of the CME has, however, turned out to be challenging because of various extrinsic effects \cite{Reis2016, Lv2021}. Moreover, since the Weyl nodes often do not reside exactly at the Fermi energy, the ballistic-limit CME is only achieved at large non-universal field strengths. 

In this work we show that in tunnel junctions between two WSMs both the CME and the Fermi arcs combine to give alternative magnetoconductance signatures of the chiral anomaly. We consider WSMs with Weyl nodes whose transverse momenta are displaced with respect to one another. Previous work has focused on the tunnel conductance across interfaces where the Fermi pockets of the two WSMs overlap \cite{Kobayashi2018,Sousa2021,Tchoumakov2021,Buccheri2022}. We instead consider non-overlapping Fermi pockets, which, in the absence of further ingredients, would simply result in a vanishing tunnel conductance. We, however, show that upon adding a magnetic field normal to the interface, the chiral Landau levels can transmit across the interface via the interface Fermi arcs, while higher Landau levels are reflected. In contrast to bulk realisations, the ballistic-limit CME across the tunnel junction (characterised by a universal, linear in field conductance), occurs around zero magnetic field, irrespective of whether the Weyl nodes are at exactly the Fermi energy. 

We further show that magnetotransport across the tunnel-junction allows for the exploration of the phenomenon of \textit{magnetic breakdown} --- magnetic-field induced quantum tunneling between disjoint equienergy contours \cite{Cohen1961,Blount1962,Shoenberg1984,Kaganov1983}. In bulk materials, level separation is typically too large for magnetic breakdown to manifest at realistic magnetic field strengths \cite{Shoenberg1984,VanDelft2018}. In a tunnel junction however, the level repulsion between Fermi arcs can be made small by weakening the coupling \cite{Sebastian2012} at the interface. The onset of magnetic breakdown causes an effective switch between topologically distinct Fermi-arc connectivities at the interface (see below), signified by a saturation of the magnetoconductance above a characteristic mangnetic-breakdown field which is controlled by the tunneling amplitude.

\textit{Tunnel-junction CME} --- 
We consider electron transport through a tunnel junction of two Weyl semimetals in the presence of a magnetic field of magnitude $B$ normal to the interface. We assume that the projection onto the interface BZ of the Fermi surfaces of all the Weyl nodes (on both sides of the interface) are separated by lattice momentum much larger than the inverse magnetic length $l^{-1}_B = \sqrt{eB/h}$. This ensures that the bulk Weyl nodes are not coupled by the applied magnetic field. 

\begin{figure}
\includegraphics[width= \columnwidth]{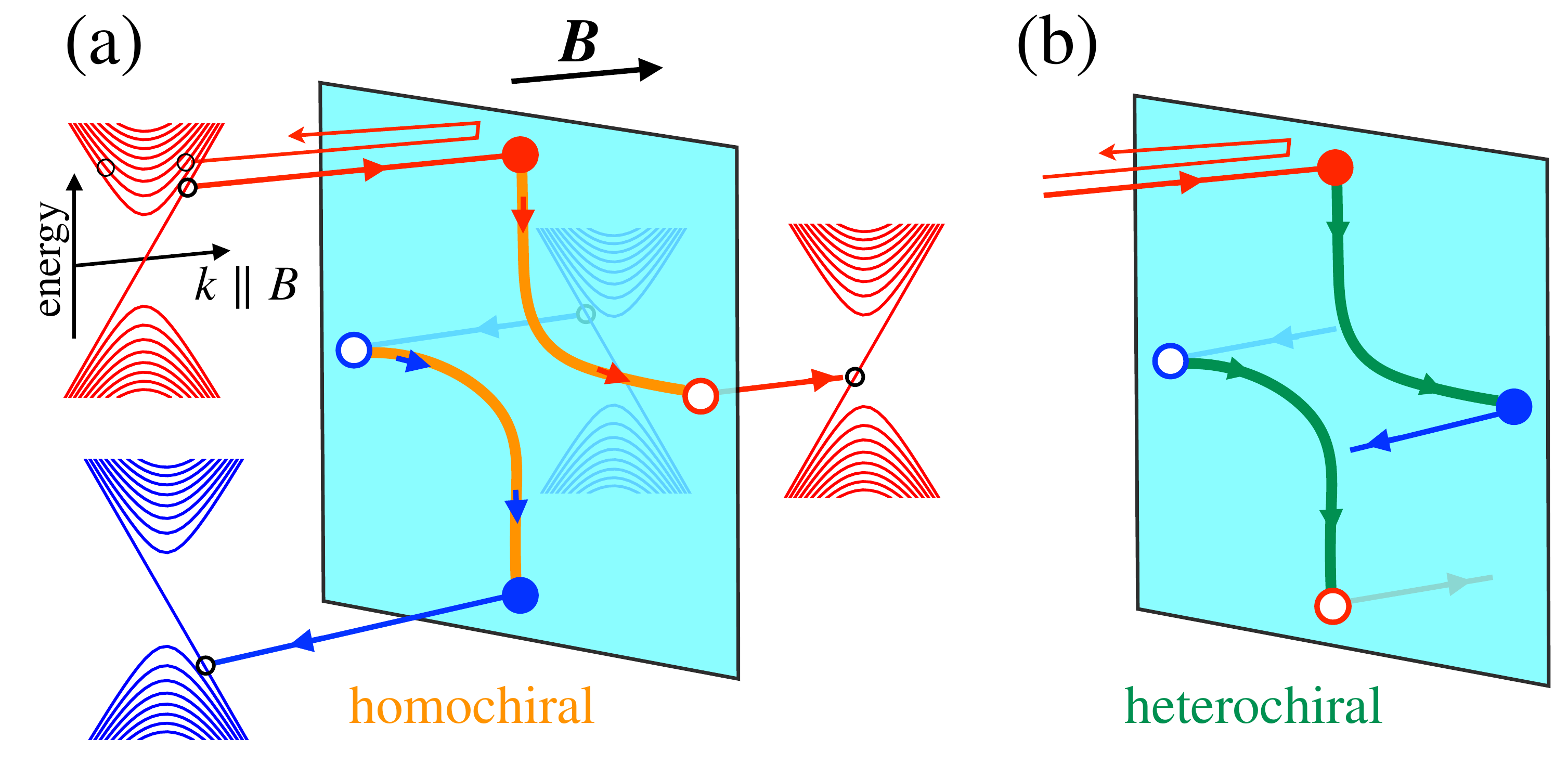}
\caption{
    Spectral flow at the interface between two Weyl semimetals with a longitudinal magnetic field (a) Interface Fermi arcs with homochiral connectivity (orange lines) lead to a full transmission of the chiral zeroth Landau level across the interface, while the higher Landau levels are reflected. (b) Fermi arcs of heterochiral connectivity (green lines) lead to reflection of all Landau levels.}
\label{fig2}
\end{figure}

In presence of the magnetic field, each Weyl node has an imbalance in the number of left and right movers because of the $N(B)$--fold degenerate chiral lowest Landau level, where $N(B)$ is the number of magnetic flux quanta through the interface.
As the spectral flow from the surplus chiral modes cannot terminate at the interface, there must exist a continuous chain of states (the interface Fermi arcs) that reconnects the interface projection of one Weyl node to that of another, as illustrated in Fig.\ \ref{fig2}. This follows from particle conservation and the observation that an infinitesimal field only couples states that are infinitesimally close in transverse momenta. 

Consequently, two such types of Fermi arcs are possible: those that connect projections of opposite-chirality Weyl nodes in the same WSM and those which connect same-chirality nodes in WSMs on opposite sides of the interface \cite{Dwivedi2018,Abdulla2021,Mathur2022,Kaushik2022}. 
We term the two connectivities \emph{heterochiral} and \emph{homochiral} connectivity, respectively. 
While the connectivity of the interface Fermi arcs is a robust topological property (See the Supplemental Material (SM) for a topological argument), their shape is non-universal and depends on system details, such as boundary potentials and the tunneling amplitude. 


We first consider the situation where the interface Fermi arcs are separated by lattice momenta much larger than $\ell_B^{-1}$. In this case, a pair of homochiral Fermi arcs perfectly transmit the incoming mode across the interface, while a pair of heterochiral Fermi arcs are totally reflected. Following a Landauer approach (see the SM for a detailed derivation), the conductance is given by 
\begin{equation}
    G = n_\mathrm{ho} N(B) \frac{e^2}{h},
\label{cond}
\end{equation}
where $n_\mathrm{ho}$ is the number of homochiral Fermi arcs and $N(B)$ is the number of incoming modes per unit area. Note that the conductance is independent of the occupation of higher Landau levels, which are pefectly reflected. This results in a universal conductance that is insensitive to material-specific details such as the energies of the Weyl nodes and velocities.

\begin{figure}
\includegraphics[width= \columnwidth]{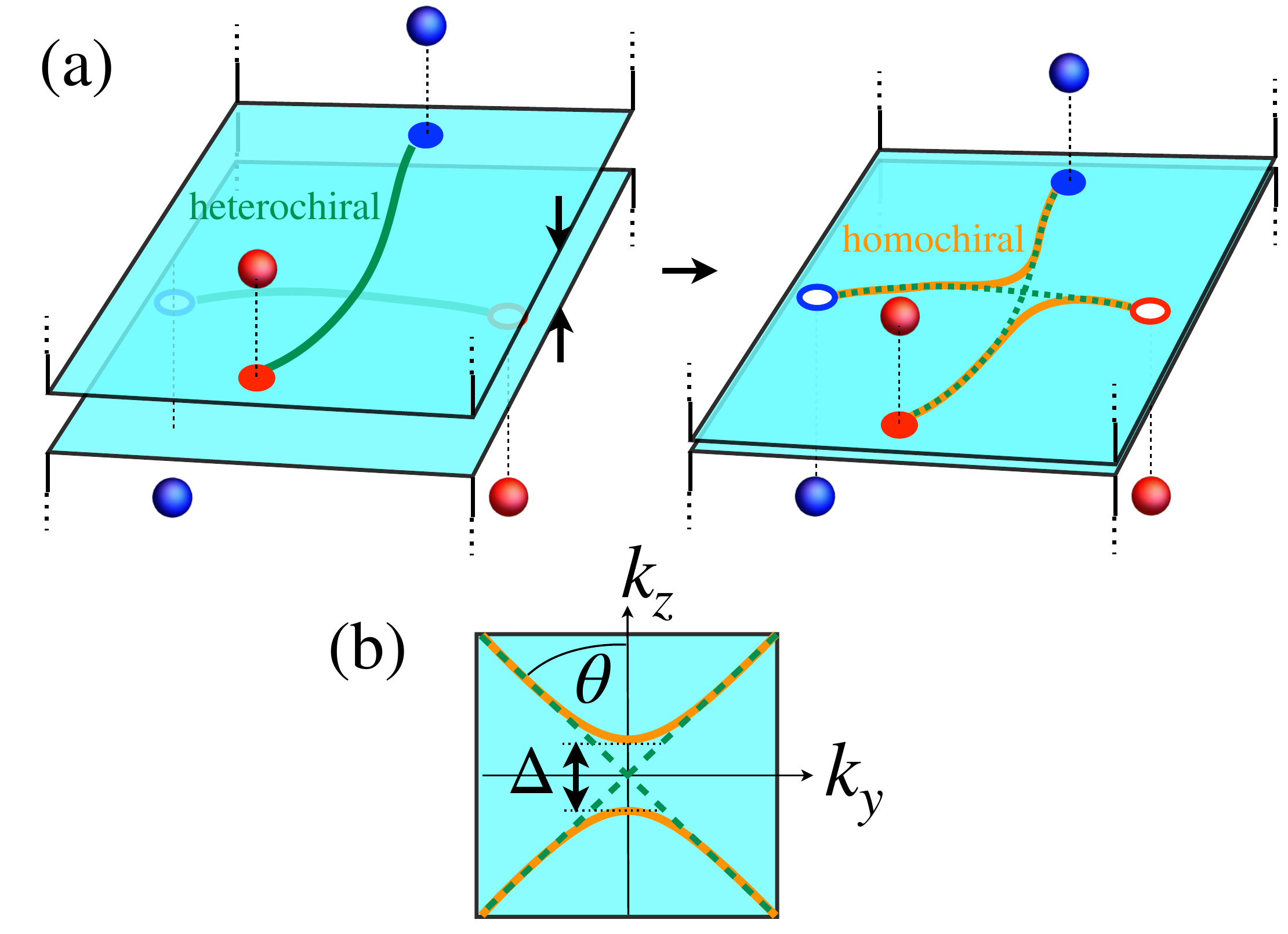}
\caption{(a) Coupling the surfaces of two WSMs with crossing Fermi arcs, whose hybridization leads to the interface Fermi arc connectivity switching from heterochiral to homochiral. 
(b) The close encounter of interface Fermi arcs for weak tunneling between the two WSMs.}
\label{fig3}
\end{figure}

\emph{Magnetic breakdown} --- 
The derivation of Eq.~\eqref{cond} breaks down once the magnetic field is large enough to enable backscattering via the interface Fermi arcs. This however, only happen once $\ell_B^{-1} \approx 0.004\,\text{\AA}^{-1} \sqrt{B[\mathrm{T}]} $ (with $B$ in Teslas) approaches the reciprocal space separation between the two chiral modes. Realistic magnetic fields can thus only couple modes whose separation is small compared to the size of the surface BZ, resulting in the phenomena of magnetic breakdown being considered a rather exotic \cite{Shoenberg1984}. However, in our setup, a close encounter of interface Fermi arcs can be achieved by a weak tunnel coupling of two WSMs whose Fermi arcs cross in the decoupled limit, as illustrated in Fig.\ \ref{fig3}(a). 


Near a close encounter of two Fermi arcs, the linearized interface Hamiltonian reads
\begin{equation}
    H_\mathrm{int}(\vk_\perp) = (v_z\Delta/2) \sigma_x
    +v_y k_y\sigma_z +v_z k_z,
    \label{hlt_int}
\end{equation}
where $\vk_\perp$ is the transverse momentum measured from the mid-point of the smallest separation between contours (see Fig.\ \ref{fig3}(b)), the Pauli matrices correspond to the two Fermi arcs from the decoupled system, the velocities $v_y$ and $v_z$ are fixed by the specific dispersion of the Fermi arcs, and $\Delta$ quantifies the hybridization strength, which depends on the tunneling amplitude. The shortest distance between the Fermi arcs is then given by $|\Delta|$ and $\theta = \tan^{-1}(v_z/v_y)$ is half the angle of the Fermi arc intersection, as shown in Fig.\ \ref{fig3}(b). 

The essential requirement for our setup is a crossing of interface Fermi arcs in the decoupled limit. At finite tunneling, such a crossing turns into a close encounter, unless protected by a symmetry deriving from the symmetries of the WSMs. Explicitly, this must forbid mass terms proportional to both $\sigma_x$ and $\sigma_y$ in Eq.~\eqref{hlt_int}. Such a protection, however, requires a lattice symmetry; which, though it may hold for interfaces between a pair of highly symmetric Weyl node configurations, would not hold for generic interfaces. The close encounter described above is thus generic so long as the decoupled Fermi arcs exhibit a crossing.  


At a finite but small $\Delta$, a longitudinal magnetic field $B$ enables quantum tunneling between the two Fermi arcs when $\ell_B^{-1}\gtrsim \Delta$.
To quantify this magnetic breakdown we deploy an analytical description following the standard formalism of Refs \cite{Cohen1961,Landau1977,Kaganov1983,Shoenberg1984}. Using the semiclassical wavefunctions of the Fermi arcs away from the encounter as scattering states that move along the arcs in accordance with the Lorentz force, we calculate the transition probability across the gap by matching the scattering states with the exact solutions of $H_\mathrm{int}$. The resulting probability of tunneling between the arcs is given by
\begin{equation}
    P = e^{-B_0/B};\qquad 
    B_0  = \frac{\pi}{4}\Delta^2 \abs{\tan \theta}.
    \label{mb}
\end{equation}
A particle passing a single encounter will thus experience a heterochiral connectivity --- and is thus reflected --- with probability $P$, so that the probability of transmission across the interface is given by $1-P$. The conductance is thus obtained by weighing Eq.~\eqref{cond} by a factor of $1-P$, so that 
\begin{equation}
    G = N(B) \frac{e^2}{h} \sum_{i=1}^{n_\mathrm{ho}}\Big(1-e^{-B_{0,i}/B}\Big),
    \label{condmb}
\end{equation}
where the sum runs over all homochiral connectivities and $B_{0,i}$ are the corresponding breakdown fields. For $B\ll B_{0,i}$ Eq.\ \eqref{cond} is recovered, while for $B\gg B_{0,i}$ the conductance saturates at $(e^2/h)N(B_{0,i})$. In the latter limit, the transmission probability approaches zero as $1/B$ (rendering the connectivity effectively heterochiral) but as $N(B)$ is linear in $B$, the conductance saturates to a constant value. 

If we further increase the magnetic field, it couples the Weyl nodes of the same chirality and opposite sides of the interfaces. In this case, the modes can transmit across the interface directly (skipping the Fermi arc), so that $N(B)$ is no longer the upper bound for the transmission, and the transmission probability becomes non-universal, depending on the number of occupied Landau levels. If the field couples opposite chirality nodes, the WSM phase is effectively destroyed \cite{Ramshaw2018}.

\emph{Lattice simulation} --- 
We test the above predictions by numerical and semi-analytical calculations on a WSM lattice model. We consider lattice models described by Hamiltonians of the form \cite{Dwivedi2016b}
\begin{align} 
  \hlt(\vk) = \hlt_x(k_x) +  \eta_y(\vk_\perp) \sigma^y + \eta_z(\vk_\perp) \sigma^z,
   \label{eq:hlt_orig} 
\end{align} 
where $\hlt_x(k_x) = \sin k_x \sigma^x + (1 - \cos k_x) \sigma^z$. The hopping strength along $x$ and the lattice constant are set to one, and $\eta_{y,z}(\vk_\perp)$ are functions of the transverse momentum $\vk_\perp \equiv (k_y, k_z)$. For $\eta_z(\vk_\perp) > -2$, the model hosts Weyl nodes in the $k_x = 0$ plane at $\vk_\perp$ satisfying $\eta_y(\vk_\perp) = \eta_z(\vk_\perp) =  0$.  
At a boundary normal to $x$, the Fermi arcs lie along $\eta_y(\vk_\perp) = 0$ for $\eta_z(\vk_\perp) < 0$ \cite{Dwivedi2016b}.

\begin{figure}[t]
\includegraphics[width=\columnwidth]{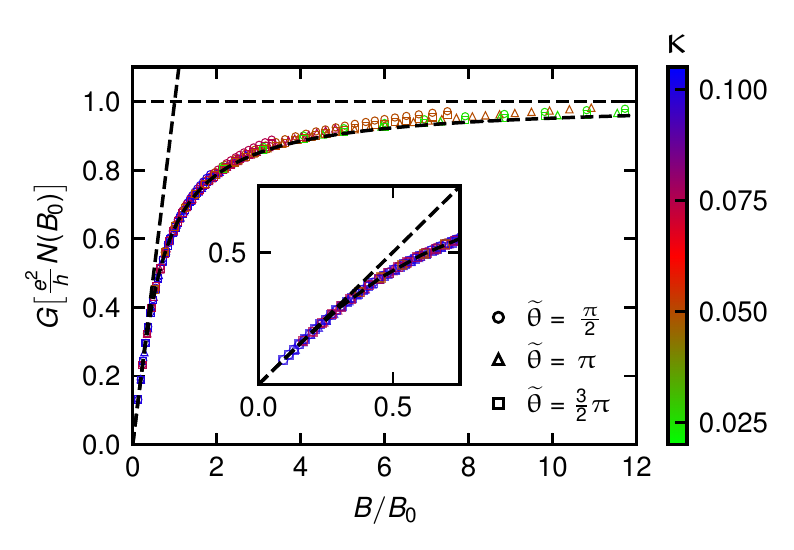}
\caption{Conductance in units of the saturation value $(e^2/h)N(B_0)$ as a function of the magnetic field in units of the breakdown field $B_0$. The numerical computation is performed for model parameters $b=\pi/2$, energy $E=0.1$ and various values of $\theta$ and $\kappa$. The analytical result (black dashed curve) fits perfectly with the numerical results for weak fields. The straight dashed lines indicate the predicted asymptotic low- and high-field behavior.
For the numerical data, the upper bound on the magnetic field is set by the smallest magnetic length $\ell_B = 3.1$ lattice units --- since smaller $\ell_B$ results in a trivial breakdown of WSM physics --- while the lower bound is set by computational limits to $13.8$ lattice units. Data points corresponding to a smaller (larger) breakdown field $B_0$ [for smaller (larger) $\kappa$] thus span a range of fields at higher (lower) values of $B/B_0$. The inset shows a close-up at small fields; highlighting the deviation from perfect transmission caused by the onset of magnetic breakdown.}
\label{fig4}
\end{figure}

We consider an interface between two WSMs with 
\begin{align}
  \eta_y^{\pm}(\vk_\perp) &= \frac1{\sin b} \left[\pm\sin b_y \sin k_y + \sin b_z \sin k_z \right], \nonumber \\ 
  \eta^{\pm}_z(\vk_\perp) &= \cos b_y + \cos b_z - \cos k_y - \cos k_z,
\label{etas}    
\end{align}
where the superscript $\pm$ refers to the left/right WSM. The corresponding Weyl nodes with  chirality $\chi$ lie at $\vk_\perp = \chi \mathbf{b}$, where
$\mathbf{b} = (b_y, b_z) = b (\sin\tht, \cos\tht)$. The tunnel junction between these two WSMs is modelled by a reduced hopping $0\leq\kappa\leq 1$ along $x$ at the interface. The conductance for this lattice model for various values of $\theta$ , computed numerically using the \emph{Kwant} package  \cite{Groth2014}, are plotted in Fig.~\ref{fig4}. 

To compare the numerical results to the analytics, we derive the  breakdown field Eq.\ \eqref{mb} from the lattice parameters. To this end, we need to determine the intersection angle $\theta$ of the Fermi arcs in the decoupled limit ($\kappa=0$) as well as the minimum separation $\Delta$ between the hybridized ($\kappa\neq 0$) Fermi arcs [cf. Fig.\ \ref{fig3}(b)]. The Fermi arcs in the decoupled limit are given by $\eta_y^{\pm}(\vk_\perp) = 0$, linearizing around $\vk_\perp=0$ gives 
\begin{equation}  
  \tan\theta = \frac{\sin b_y}{\sin b_z}
             = \frac{\sin (b\sin\tht)}{\sin (b\cos\tht)}.
  \label{th}
\end{equation} 
We next derive  the interface Fermi arcs at a finite coupling $\kappa$ using a transfer-matrix approach \cite{Dwivedi2016, Dwivedi2016b}, as described in the SM. To leading order in $\kappa \ll 1$, the minimum separation is given by
\begin{equation}
  \Delta  = 2 \kappa \beta(2+\beta) \frac{\sin b}{\sin b_y }, \label{del}
\end{equation}
where $\beta = \cos b_y + \cos b_z - 2$. The breakdown field can then be determined by Eqs.\ \eqref{mb}, \eqref{th}, and \eqref{del}. The conductance computed using \eqref{condmb} with the analytically computed breakdown field shows excellent agreement with the numerical results so long as the magnetic length is much larger than the lattice spacing, as shown in  Fig.~\ref{fig4}. 



\emph{Discussion and conclusion} --- 
In this article, we consider the magnetoconductance across a tunnel barrier between two WSMs arranged such that the projection of the Weyl node's Fermi surfaces onto the interface BZ are well separated and thereby aren't coupled by a magnetic field. At generic couplings of the WSMs the interface Fermi arcs, which come in two topologically distinct connectivity types (heterochiral and homochiral), will typically also be well separated. In this regime the system displays a universal tunnel magnetoconductance of $(e^2/h)\, N(B)$ (where $N(B)$ is the number of flux quanta through the interface) for each pair of homochiral-connectivity Fermi arcs. While the CME of a bulk WSM in the ballistic limit is also characterised by a universal magnetoconductance where the number of Fermi-arc pairs is replaced by the number of Weyl-node pairs, it requires a minimal magnetic-field strength, whose scale is set by the energy of the Weyl nodes and by diffusion properties \cite{Altland2016, Burkov2017}. In contrast, the tunnel conductance considered here is independent of such details because the interface is intransparent for higher Landau level and thus acts like a filter. An alternative bulk system showing a similar effect of the ballistic-limit CME extending down to zero magnetic field independent of system details is achieved in Fermi-arc metals \cite{Breitkreiz2022}, recently predicted by Brouwer and one of us.


Such interfaces can also be used to realize the phenomena of magnetic breakdown. This requires a close encounter of two Fermi arcs so as to enable magnetic field induced quantum tunneling between arcs. Such an encounter is generic between WSMs whose interface Fermi arcs cross in the decoupled limit. The magnetic breakdown leads to a suppression of the transmission probability for field strengths above a characteristic breakdown field $B_0$ (set by the coupling strength). Since the magnetic breakdown effectively turns a homochiral connectivity into a heterochiral connectivity, one might expect the conductance to drop to zero above $B_0$. The increased probability of experiencing a heterochiral connectivity at a higher field is however balanced by the increased degeneracy of transmitted modes, thus leading to the saturation of the conductance at a finite value of $(e^2/h)\, N(B_0)$. 

\textit{Acknowledgments}. 
We thank Piet W.\ Brouwer, Achim Rosch, Reinhold Egger, and Francesco Buccheri for useful discussions. This research was supported by projects A02 and A03 of the CRC-TR 183 “Entangled States of Matter” and Grant No.\ 18688556 of the Deutsche Forschungsgemeinschaft (DFG, German Science Foundation).

\bibliography{library}	

\begin{thebibliography}{39}%
\makeatletter
\providecommand \@ifxundefined [1]{%
 \@ifx{#1\undefined}
}%
\providecommand \@ifnum [1]{%
 \ifnum #1\expandafter \@firstoftwo
 \else \expandafter \@secondoftwo
 \fi
}%
\providecommand \@ifx [1]{%
 \ifx #1\expandafter \@firstoftwo
 \else \expandafter \@secondoftwo
 \fi
}%
\providecommand \natexlab [1]{#1}%
\providecommand \enquote  [1]{``#1''}%
\providecommand \bibnamefont  [1]{#1}%
\providecommand \bibfnamefont [1]{#1}%
\providecommand \citenamefont [1]{#1}%
\providecommand \href@noop [0]{\@secondoftwo}%
\providecommand \href [0]{\begingroup \@sanitize@url \@href}%
\providecommand \@href[1]{\@@startlink{#1}\@@href}%
\providecommand \@@href[1]{\endgroup#1\@@endlink}%
\providecommand \@sanitize@url [0]{\catcode `\\12\catcode `\$12\catcode
  `\&12\catcode `\#12\catcode `\^12\catcode `\_12\catcode `\%12\relax}%
\providecommand \@@startlink[1]{}%
\providecommand \@@endlink[0]{}%
\providecommand \url  [0]{\begingroup\@sanitize@url \@url }%
\providecommand \@url [1]{\endgroup\@href {#1}{\urlprefix }}%
\providecommand \urlprefix  [0]{URL }%
\providecommand \Eprint [0]{\href }%
\providecommand \doibase [0]{http://dx.doi.org/}%
\providecommand \selectlanguage [0]{\@gobble}%
\providecommand \bibinfo  [0]{\@secondoftwo}%
\providecommand \bibfield  [0]{\@secondoftwo}%
\providecommand \translation [1]{[#1]}%
\providecommand \BibitemOpen [0]{}%
\providecommand \bibitemStop [0]{}%
\providecommand \bibitemNoStop [0]{.\EOS\space}%
\providecommand \EOS [0]{\spacefactor3000\relax}%
\providecommand \BibitemShut  [1]{\csname bibitem#1\endcsname}%
\let\auto@bib@innerbib\@empty
\bibitem [{\citenamefont {Wan}\ \emph {et~al.}(2011)\citenamefont {Wan},
  \citenamefont {Turner}, \citenamefont {Vishwanath},\ and\ \citenamefont
  {Savrasov}}]{Wan2011}%
  \BibitemOpen
  \bibfield  {author} {\bibinfo {author} {\bibfnamefont {X.}~\bibnamefont
  {Wan}}, \bibinfo {author} {\bibfnamefont {A.~M.}\ \bibnamefont {Turner}},
  \bibinfo {author} {\bibfnamefont {A.}~\bibnamefont {Vishwanath}}, \ and\
  \bibinfo {author} {\bibfnamefont {S.~Y.}\ \bibnamefont {Savrasov}},\ }\href
  {\doibase 10.1103/PhysRevB.83.205101} {\bibfield  {journal} {\bibinfo
  {journal} {Phys. Rev. B}\ }\textbf {\bibinfo {volume} {83}},\ \bibinfo
  {pages} {205101} (\bibinfo {year} {2011})}\BibitemShut {NoStop}%
\bibitem [{\citenamefont {Burkov}\ and\ \citenamefont
  {Balents}(2011)}]{Burkov2011}%
  \BibitemOpen
  \bibfield  {author} {\bibinfo {author} {\bibfnamefont {A.~A.}\ \bibnamefont
  {Burkov}}\ and\ \bibinfo {author} {\bibfnamefont {L.}~\bibnamefont
  {Balents}},\ }\href {\doibase https://doi.org/10.1103/PhysRevLett.107.127205}
  {\bibfield  {journal} {\bibinfo  {journal} {Phys. Rev. Lett}\ }\textbf
  {\bibinfo {volume} {107}},\ \bibinfo {pages} {127205} (\bibinfo {year}
  {2011})}\BibitemShut {NoStop}%
\bibitem [{\citenamefont {Xu}\ \emph {et~al.}(2011)\citenamefont {Xu},
  \citenamefont {Weng}, \citenamefont {Wang}, \citenamefont {Dai},\ and\
  \citenamefont {Fang}}]{Xu2011}%
  \BibitemOpen
  \bibfield  {author} {\bibinfo {author} {\bibfnamefont {G.}~\bibnamefont
  {Xu}}, \bibinfo {author} {\bibfnamefont {H.}~\bibnamefont {Weng}}, \bibinfo
  {author} {\bibfnamefont {Z.}~\bibnamefont {Wang}}, \bibinfo {author}
  {\bibfnamefont {X.}~\bibnamefont {Dai}}, \ and\ \bibinfo {author}
  {\bibfnamefont {Z.}~\bibnamefont {Fang}},\ }\href {\doibase
  10.1103/PhysRevLett.107.186806} {\bibfield  {journal} {\bibinfo  {journal}
  {Phys. Rev. Lett.}\ }\textbf {\bibinfo {volume} {107}},\ \bibinfo {pages}
  {186806} (\bibinfo {year} {2011})}\BibitemShut {NoStop}%
\bibitem [{\citenamefont {Xu}\ \emph {et~al.}(2015{\natexlab{a}})\citenamefont
  {Xu}, \citenamefont {Liu}, \citenamefont {Kushwaha}, \citenamefont {Sankar},
  \citenamefont {Krizan}, \citenamefont {Belopolski}, \citenamefont {Neupane},
  \citenamefont {Bian}, \citenamefont {Alidoust}, \citenamefont {Chang},
  \citenamefont {Jeng}, \citenamefont {Huang}, \citenamefont {Tsai},
  \citenamefont {Lin}, \citenamefont {Shibayev}, \citenamefont {Chou},
  \citenamefont {Cava},\ and\ \citenamefont {Hasan}}]{Xu2015}%
  \BibitemOpen
  \bibfield  {author} {\bibinfo {author} {\bibfnamefont {S.-Y.}\ \bibnamefont
  {Xu}}, \bibinfo {author} {\bibfnamefont {C.}~\bibnamefont {Liu}}, \bibinfo
  {author} {\bibfnamefont {S.~K.}\ \bibnamefont {Kushwaha}}, \bibinfo {author}
  {\bibfnamefont {R.}~\bibnamefont {Sankar}}, \bibinfo {author} {\bibfnamefont
  {J.~W.}\ \bibnamefont {Krizan}}, \bibinfo {author} {\bibfnamefont
  {I.}~\bibnamefont {Belopolski}}, \bibinfo {author} {\bibfnamefont
  {M.}~\bibnamefont {Neupane}}, \bibinfo {author} {\bibfnamefont
  {G.}~\bibnamefont {Bian}}, \bibinfo {author} {\bibfnamefont {N.}~\bibnamefont
  {Alidoust}}, \bibinfo {author} {\bibfnamefont {T.-R.}\ \bibnamefont {Chang}},
  \bibinfo {author} {\bibfnamefont {H.-T.}\ \bibnamefont {Jeng}}, \bibinfo
  {author} {\bibfnamefont {C.-Y.}\ \bibnamefont {Huang}}, \bibinfo {author}
  {\bibfnamefont {W.-F.}\ \bibnamefont {Tsai}}, \bibinfo {author}
  {\bibfnamefont {H.}~\bibnamefont {Lin}}, \bibinfo {author} {\bibfnamefont
  {P.~P.}\ \bibnamefont {Shibayev}}, \bibinfo {author} {\bibfnamefont {F.-C.}\
  \bibnamefont {Chou}}, \bibinfo {author} {\bibfnamefont {R.~J.}\ \bibnamefont
  {Cava}}, \ and\ \bibinfo {author} {\bibfnamefont {M.~Z.}\ \bibnamefont
  {Hasan}},\ }\href {\doibase DOI: 10.1126/science.1256742} {\bibfield
  {journal} {\bibinfo  {journal} {Science}\ }\textbf {\bibinfo {volume}
  {347}},\ \bibinfo {pages} {294} (\bibinfo {year}
  {2015}{\natexlab{a}})}\BibitemShut {NoStop}%
\bibitem [{\citenamefont {Xu}\ \emph {et~al.}(2015{\natexlab{b}})\citenamefont
  {Xu}, \citenamefont {Alidoust}, \citenamefont {Belopolski}, \citenamefont
  {Yuan}, \citenamefont {Bian}, \citenamefont {Chang}, \citenamefont {Zheng},
  \citenamefont {Strocov}, \citenamefont {Sanchez}, \citenamefont {Chang},
  \citenamefont {Zhang}, \citenamefont {Mou}, \citenamefont {Wu}, \citenamefont
  {Huang}, \citenamefont {Lee}, \citenamefont {Huang}, \citenamefont {Wang},
  \citenamefont {Bansil}, \citenamefont {Jeng}, \citenamefont {Neupert},
  \citenamefont {Kaminski}, \citenamefont {Lin}, \citenamefont {Jia},\ and\
  \citenamefont {{Zahid Hasan}}}]{Xu2015b}%
  \BibitemOpen
  \bibfield  {author} {\bibinfo {author} {\bibfnamefont {S.-Y.}\ \bibnamefont
  {Xu}}, \bibinfo {author} {\bibfnamefont {N.}~\bibnamefont {Alidoust}},
  \bibinfo {author} {\bibfnamefont {I.}~\bibnamefont {Belopolski}}, \bibinfo
  {author} {\bibfnamefont {Z.}~\bibnamefont {Yuan}}, \bibinfo {author}
  {\bibfnamefont {G.}~\bibnamefont {Bian}}, \bibinfo {author} {\bibfnamefont
  {T.-R.}\ \bibnamefont {Chang}}, \bibinfo {author} {\bibfnamefont
  {H.}~\bibnamefont {Zheng}}, \bibinfo {author} {\bibfnamefont {V.~N.}\
  \bibnamefont {Strocov}}, \bibinfo {author} {\bibfnamefont {D.~S.}\
  \bibnamefont {Sanchez}}, \bibinfo {author} {\bibfnamefont {G.}~\bibnamefont
  {Chang}}, \bibinfo {author} {\bibfnamefont {C.}~\bibnamefont {Zhang}},
  \bibinfo {author} {\bibfnamefont {D.}~\bibnamefont {Mou}}, \bibinfo {author}
  {\bibfnamefont {Y.}~\bibnamefont {Wu}}, \bibinfo {author} {\bibfnamefont
  {L.}~\bibnamefont {Huang}}, \bibinfo {author} {\bibfnamefont {C.-C.}\
  \bibnamefont {Lee}}, \bibinfo {author} {\bibfnamefont {S.-M.}\ \bibnamefont
  {Huang}}, \bibinfo {author} {\bibfnamefont {B.}~\bibnamefont {Wang}},
  \bibinfo {author} {\bibfnamefont {A.}~\bibnamefont {Bansil}}, \bibinfo
  {author} {\bibfnamefont {H.-T.}\ \bibnamefont {Jeng}}, \bibinfo {author}
  {\bibfnamefont {T.}~\bibnamefont {Neupert}}, \bibinfo {author} {\bibfnamefont
  {A.}~\bibnamefont {Kaminski}}, \bibinfo {author} {\bibfnamefont
  {H.}~\bibnamefont {Lin}}, \bibinfo {author} {\bibfnamefont {S.}~\bibnamefont
  {Jia}}, \ and\ \bibinfo {author} {\bibfnamefont {M.}~\bibnamefont {{Zahid
  Hasan}}},\ }\href {\doibase 10.1038/nphys3437} {\bibfield  {journal}
  {\bibinfo  {journal} {Nat. Phys.}\ }\textbf {\bibinfo {volume} {11}},\
  \bibinfo {pages} {748} (\bibinfo {year} {2015}{\natexlab{b}})}\BibitemShut
  {NoStop}%
\bibitem [{\citenamefont {Lv}\ \emph {et~al.}(2015)\citenamefont {Lv},
  \citenamefont {Weng}, \citenamefont {Fu}, \citenamefont {Wang}, \citenamefont
  {Miao}, \citenamefont {Ma}, \citenamefont {Richard}, \citenamefont {Huang},
  \citenamefont {Zhao}, \citenamefont {Chen}, \citenamefont {Fang},
  \citenamefont {Dai}, \citenamefont {Qian},\ and\ \citenamefont
  {Ding}}]{Lv2015}%
  \BibitemOpen
  \bibfield  {author} {\bibinfo {author} {\bibfnamefont {B.~Q.}\ \bibnamefont
  {Lv}}, \bibinfo {author} {\bibfnamefont {H.~M.}\ \bibnamefont {Weng}},
  \bibinfo {author} {\bibfnamefont {B.~B.}\ \bibnamefont {Fu}}, \bibinfo
  {author} {\bibfnamefont {X.~P.}\ \bibnamefont {Wang}}, \bibinfo {author}
  {\bibfnamefont {H.}~\bibnamefont {Miao}}, \bibinfo {author} {\bibfnamefont
  {J.}~\bibnamefont {Ma}}, \bibinfo {author} {\bibfnamefont {P.}~\bibnamefont
  {Richard}}, \bibinfo {author} {\bibfnamefont {X.~C.}\ \bibnamefont {Huang}},
  \bibinfo {author} {\bibfnamefont {L.~X.}\ \bibnamefont {Zhao}}, \bibinfo
  {author} {\bibfnamefont {G.~F.}\ \bibnamefont {Chen}}, \bibinfo {author}
  {\bibfnamefont {Z.}~\bibnamefont {Fang}}, \bibinfo {author} {\bibfnamefont
  {X.}~\bibnamefont {Dai}}, \bibinfo {author} {\bibfnamefont {T.}~\bibnamefont
  {Qian}}, \ and\ \bibinfo {author} {\bibfnamefont {H.}~\bibnamefont {Ding}},\
  }\href {\doibase 10.1103/PhysRevX.5.031013} {\bibfield  {journal} {\bibinfo
  {journal} {Phys. Rev. X}\ }\textbf {\bibinfo {volume} {5}},\ \bibinfo {pages}
  {031013} (\bibinfo {year} {2015})}\BibitemShut {NoStop}%
\bibitem [{\citenamefont {Armitage}\ \emph {et~al.}(2018)\citenamefont
  {Armitage}, \citenamefont {Mele},\ and\ \citenamefont
  {Vishwanath}}]{Armitage2017}%
  \BibitemOpen
  \bibfield  {author} {\bibinfo {author} {\bibfnamefont {N.~P.}\ \bibnamefont
  {Armitage}}, \bibinfo {author} {\bibfnamefont {E.~J.}\ \bibnamefont {Mele}},
  \ and\ \bibinfo {author} {\bibfnamefont {A.}~\bibnamefont {Vishwanath}},\
  }\href {\doibase 10.1103/RevModPhys.90.015001} {\bibfield  {journal}
  {\bibinfo  {journal} {Rev. Mod. Phys.}\ }\textbf {\bibinfo {volume} {90}},\
  \bibinfo {pages} {015001} (\bibinfo {year} {2018})}\BibitemShut {NoStop}%
\bibitem [{\citenamefont {Yan}\ and\ \citenamefont {Felser}(2017)}]{Yan2017}%
  \BibitemOpen
  \bibfield  {author} {\bibinfo {author} {\bibfnamefont {B.}~\bibnamefont
  {Yan}}\ and\ \bibinfo {author} {\bibfnamefont {C.}~\bibnamefont {Felser}},\
  }\href {\doibase 10.1146/annurev-conmatphys-031016-025458} {\bibfield
  {journal} {\bibinfo  {journal} {Annu. Rev. Condens. Matter Phys.}\ }\textbf
  {\bibinfo {volume} {8}},\ \bibinfo {pages} {337} (\bibinfo {year}
  {2017})}\BibitemShut {NoStop}%
\bibitem [{\citenamefont {Burkov}(2018)}]{Burkov2017}%
  \BibitemOpen
  \bibfield  {author} {\bibinfo {author} {\bibfnamefont {A.~A.}\ \bibnamefont
  {Burkov}},\ }\href {\doibase 10.1146/annurev-conmatphys-033117-054129}
  {\bibfield  {journal} {\bibinfo  {journal} {Annu. Rev. Condens. Matter
  Phys.}\ }\textbf {\bibinfo {volume} {9}},\ \bibinfo {pages} {359} (\bibinfo
  {year} {2018})}\BibitemShut {NoStop}%
\bibitem [{\citenamefont {Hasan}\ \emph {et~al.}(2021)\citenamefont {Hasan},
  \citenamefont {Chang}, \citenamefont {Belopolski}, \citenamefont {Bian},
  \citenamefont {Xu},\ and\ \citenamefont {Yin}}]{Hasan2021}%
  \BibitemOpen
  \bibfield  {author} {\bibinfo {author} {\bibfnamefont {M.~Z.}\ \bibnamefont
  {Hasan}}, \bibinfo {author} {\bibfnamefont {G.}~\bibnamefont {Chang}},
  \bibinfo {author} {\bibfnamefont {I.}~\bibnamefont {Belopolski}}, \bibinfo
  {author} {\bibfnamefont {G.}~\bibnamefont {Bian}}, \bibinfo {author}
  {\bibfnamefont {S.~Y.}\ \bibnamefont {Xu}}, \ and\ \bibinfo {author}
  {\bibfnamefont {J.~X.}\ \bibnamefont {Yin}},\ }\href {\doibase
  10.1038/s41578-021-00301-3} {\bibfield  {journal} {\bibinfo  {journal} {Nat.
  Rev. Mater.}\ }\textbf {\bibinfo {volume} {6}},\ \bibinfo {pages} {784}
  (\bibinfo {year} {2021})}\BibitemShut {NoStop}%
\bibitem [{\citenamefont {Bernevig}\ \emph {et~al.}(2022)\citenamefont
  {Bernevig}, \citenamefont {Felser},\ and\ \citenamefont
  {Beidenkopf}}]{Bernevig2022a}%
  \BibitemOpen
  \bibfield  {author} {\bibinfo {author} {\bibfnamefont {B.~A.}\ \bibnamefont
  {Bernevig}}, \bibinfo {author} {\bibfnamefont {C.}~\bibnamefont {Felser}}, \
  and\ \bibinfo {author} {\bibfnamefont {H.}~\bibnamefont {Beidenkopf}},\
  }\href {\doibase 10.1038/s41586-021-04105-x} {\bibfield  {journal} {\bibinfo
  {journal} {Nature}\ }\textbf {\bibinfo {volume} {603}},\ \bibinfo {pages}
  {41} (\bibinfo {year} {2022})}\BibitemShut {NoStop}%
\bibitem [{\citenamefont {Adler}(1969)}]{Adler1969}%
  \BibitemOpen
  \bibfield  {author} {\bibinfo {author} {\bibfnamefont {S.~L.}\ \bibnamefont
  {Adler}},\ }\href@noop {} {\bibfield  {journal} {\bibinfo  {journal} {Phys.
  Rev.}\ }\textbf {\bibinfo {volume} {177(5)}},\ \bibinfo {pages} {2426}
  (\bibinfo {year} {1969})}\BibitemShut {NoStop}%
\bibitem [{\citenamefont {Bell}\ and\ \citenamefont {Jackiw}(1969)}]{Bell1969}%
  \BibitemOpen
  \bibfield  {author} {\bibinfo {author} {\bibfnamefont {J.~S.}\ \bibnamefont
  {Bell}}\ and\ \bibinfo {author} {\bibfnamefont {R.~W.}\ \bibnamefont
  {Jackiw}},\ }\href@noop {} {\bibfield  {journal} {\bibinfo  {journal} {Nuovo
  Cim.}\ }\textbf {\bibinfo {volume} {60}},\ \bibinfo {pages} {47} (\bibinfo
  {year} {1969})}\BibitemShut {NoStop}%
\bibitem [{\citenamefont {Nielsen}\ and\ \citenamefont
  {Ninomiya}(1983)}]{Nielsen1983}%
  \BibitemOpen
  \bibfield  {author} {\bibinfo {author} {\bibfnamefont {H.~B.}\ \bibnamefont
  {Nielsen}}\ and\ \bibinfo {author} {\bibfnamefont {M.}~\bibnamefont
  {Ninomiya}},\ }\href@noop {} {\bibfield  {journal} {\bibinfo  {journal}
  {Phys. Lett. B}\ }\textbf {\bibinfo {volume} {130(6)}},\ \bibinfo {pages}
  {389} (\bibinfo {year} {1983})}\BibitemShut {NoStop}%
\bibitem [{\citenamefont {Son}\ and\ \citenamefont {Spivak}(2013)}]{Son2013a}%
  \BibitemOpen
  \bibfield  {author} {\bibinfo {author} {\bibfnamefont {D.~T.}\ \bibnamefont
  {Son}}\ and\ \bibinfo {author} {\bibfnamefont {B.~Z.}\ \bibnamefont
  {Spivak}},\ }\href {\doibase 10.1103/PhysRevB.88.104412} {\bibfield
  {journal} {\bibinfo  {journal} {Phys. Rev. B}\ }\textbf {\bibinfo {volume}
  {88}},\ \bibinfo {pages} {104412} (\bibinfo {year} {2013})}\BibitemShut
  {NoStop}%
\bibitem [{\citenamefont {Xiong}\ \emph {et~al.}(2015)\citenamefont {Xiong},
  \citenamefont {Kushwaha}, \citenamefont {Liang}, \citenamefont {Krizan},
  \citenamefont {Hirschberger}, \citenamefont {Wang}, \citenamefont {Cava},\
  and\ \citenamefont {Ong}}]{Xiong2015a}%
  \BibitemOpen
  \bibfield  {author} {\bibinfo {author} {\bibfnamefont {J.}~\bibnamefont
  {Xiong}}, \bibinfo {author} {\bibfnamefont {S.~K.}\ \bibnamefont {Kushwaha}},
  \bibinfo {author} {\bibfnamefont {T.}~\bibnamefont {Liang}}, \bibinfo
  {author} {\bibfnamefont {J.~W.}\ \bibnamefont {Krizan}}, \bibinfo {author}
  {\bibfnamefont {M.}~\bibnamefont {Hirschberger}}, \bibinfo {author}
  {\bibfnamefont {W.}~\bibnamefont {Wang}}, \bibinfo {author} {\bibfnamefont
  {R.~J.}\ \bibnamefont {Cava}}, \ and\ \bibinfo {author} {\bibfnamefont
  {N.~P.}\ \bibnamefont {Ong}},\ }\href {\doibase 10.1126/science.aac6089}
  {\bibfield  {journal} {\bibinfo  {journal} {Science}\ }\textbf {\bibinfo
  {volume} {350}},\ \bibinfo {pages} {413} (\bibinfo {year}
  {2015})}\BibitemShut {NoStop}%
\bibitem [{\citenamefont {Altland}\ and\ \citenamefont
  {Bagrets}(2016)}]{Altland2016}%
  \BibitemOpen
  \bibfield  {author} {\bibinfo {author} {\bibfnamefont {A.}~\bibnamefont
  {Altland}}\ and\ \bibinfo {author} {\bibfnamefont {D.}~\bibnamefont
  {Bagrets}},\ }\href {\doibase 10.1103/PhysRevB.93.075113} {\bibfield
  {journal} {\bibinfo  {journal} {Phys. Rev. B}\ }\textbf {\bibinfo {volume}
  {93}},\ \bibinfo {pages} {75113} (\bibinfo {year} {2016})}\BibitemShut
  {NoStop}%
\bibitem [{\citenamefont {dos Reis}\ \emph {et~al.}(2016)\citenamefont {dos
  Reis}, \citenamefont {Ajeesh}, \citenamefont {Kumar}, \citenamefont {Arnold},
  \citenamefont {Shekhar}, \citenamefont {Naumann}, \citenamefont {Schmidt},
  \citenamefont {Nicklas},\ and\ \citenamefont {Hassinger}}]{Reis2016}%
  \BibitemOpen
  \bibfield  {author} {\bibinfo {author} {\bibfnamefont {R.~D.}\ \bibnamefont
  {dos Reis}}, \bibinfo {author} {\bibfnamefont {M.~O.}\ \bibnamefont
  {Ajeesh}}, \bibinfo {author} {\bibfnamefont {N.}~\bibnamefont {Kumar}},
  \bibinfo {author} {\bibfnamefont {F.}~\bibnamefont {Arnold}}, \bibinfo
  {author} {\bibfnamefont {C.}~\bibnamefont {Shekhar}}, \bibinfo {author}
  {\bibfnamefont {M.}~\bibnamefont {Naumann}}, \bibinfo {author} {\bibfnamefont
  {M.}~\bibnamefont {Schmidt}}, \bibinfo {author} {\bibfnamefont
  {M.}~\bibnamefont {Nicklas}}, \ and\ \bibinfo {author} {\bibfnamefont
  {E.}~\bibnamefont {Hassinger}},\ }\href {\doibase
  10.1088/1367-2630/18/8/085006} {\bibfield  {journal} {\bibinfo  {journal}
  {New J. Phys.}\ }\textbf {\bibinfo {volume} {18}},\ \bibinfo {pages} {085006}
  (\bibinfo {year} {2016})}\BibitemShut {NoStop}%
\bibitem [{\citenamefont {Lv}\ \emph {et~al.}(2021)\citenamefont {Lv},
  \citenamefont {Qian},\ and\ \citenamefont {Ding}}]{Lv2021}%
  \BibitemOpen
  \bibfield  {author} {\bibinfo {author} {\bibfnamefont {B.~Q.}\ \bibnamefont
  {Lv}}, \bibinfo {author} {\bibfnamefont {T.}~\bibnamefont {Qian}}, \ and\
  \bibinfo {author} {\bibfnamefont {H.}~\bibnamefont {Ding}},\ }\href {\doibase
  10.1103/RevModPhys.93.025002} {\bibfield  {journal} {\bibinfo  {journal}
  {Rev. Mod. Phys.}\ }\textbf {\bibinfo {volume} {93}},\ \bibinfo {pages}
  {25002} (\bibinfo {year} {2021})}\BibitemShut {NoStop}%
\bibitem [{\citenamefont {Kobayashi}\ \emph {et~al.}(2018)\citenamefont
  {Kobayashi}, \citenamefont {Ominato},\ and\ \citenamefont
  {Nomura}}]{Kobayashi2018}%
  \BibitemOpen
  \bibfield  {author} {\bibinfo {author} {\bibfnamefont {K.}~\bibnamefont
  {Kobayashi}}, \bibinfo {author} {\bibfnamefont {Y.}~\bibnamefont {Ominato}},
  \ and\ \bibinfo {author} {\bibfnamefont {K.}~\bibnamefont {Nomura}},\ }\href
  {\doibase 10.7566/JPSJ.87.073707} {\bibfield  {journal} {\bibinfo  {journal}
  {J. Phys. Soc. Japan}\ }\textbf {\bibinfo {volume} {87}},\ \bibinfo {pages}
  {73707} (\bibinfo {year} {2018})}\BibitemShut {NoStop}%
\bibitem [{\citenamefont {Sousa}\ \emph {et~al.}(2021)\citenamefont {Sousa},
  \citenamefont {Ascencio}, \citenamefont {Haney}, \citenamefont {Wang},\ and\
  \citenamefont {Low}}]{Sousa2021}%
  \BibitemOpen
  \bibfield  {author} {\bibinfo {author} {\bibfnamefont {D.~J. P.~D.}\
  \bibnamefont {Sousa}}, \bibinfo {author} {\bibfnamefont {C.~O.}\ \bibnamefont
  {Ascencio}}, \bibinfo {author} {\bibfnamefont {P.~M.}\ \bibnamefont {Haney}},
  \bibinfo {author} {\bibfnamefont {J.~P.}\ \bibnamefont {Wang}}, \ and\
  \bibinfo {author} {\bibfnamefont {T.}~\bibnamefont {Low}},\ }\href {\doibase
  10.1103/PhysRevB.104.L041401} {\bibfield  {journal} {\bibinfo  {journal}
  {Phys. Rev. B}\ }\textbf {\bibinfo {volume} {104}},\ \bibinfo {pages}
  {L041401} (\bibinfo {year} {2021})}\BibitemShut {NoStop}%
\bibitem [{\citenamefont {Tchoumakov}\ \emph {et~al.}(2021)\citenamefont
  {Tchoumakov}, \citenamefont {Bujnowski}, \citenamefont {Noky}, \citenamefont
  {Gooth}, \citenamefont {Grushin},\ and\ \citenamefont
  {Cayssol}}]{Tchoumakov2021}%
  \BibitemOpen
  \bibfield  {author} {\bibinfo {author} {\bibfnamefont {S.}~\bibnamefont
  {Tchoumakov}}, \bibinfo {author} {\bibfnamefont {B.}~\bibnamefont
  {Bujnowski}}, \bibinfo {author} {\bibfnamefont {J.}~\bibnamefont {Noky}},
  \bibinfo {author} {\bibfnamefont {J.}~\bibnamefont {Gooth}}, \bibinfo
  {author} {\bibfnamefont {A.~G.}\ \bibnamefont {Grushin}}, \ and\ \bibinfo
  {author} {\bibfnamefont {J.}~\bibnamefont {Cayssol}},\ }\href {\doibase
  10.1103/PhysRevB.104.125308} {\bibfield  {journal} {\bibinfo  {journal}
  {Phys. Rev. B}\ }\textbf {\bibinfo {volume} {104}},\ \bibinfo {pages} {1}
  (\bibinfo {year} {2021})}\BibitemShut {NoStop}%
\bibitem [{\citenamefont {Buccheri}\ \emph {et~al.}(2022)\citenamefont
  {Buccheri}, \citenamefont {Egger},\ and\ \citenamefont {{De
  Martino}}}]{Buccheri2022}%
  \BibitemOpen
  \bibfield  {author} {\bibinfo {author} {\bibfnamefont {F.}~\bibnamefont
  {Buccheri}}, \bibinfo {author} {\bibfnamefont {R.}~\bibnamefont {Egger}}, \
  and\ \bibinfo {author} {\bibfnamefont {A.}~\bibnamefont {{De Martino}}},\
  }\href {\doibase 10.1103/PhysRevB.106.045413} {\bibfield  {journal} {\bibinfo
   {journal} {Phys. Rev. B}\ }\textbf {\bibinfo {volume} {106}},\ \bibinfo
  {pages} {45413} (\bibinfo {year} {2022})}\BibitemShut {NoStop}%
\bibitem [{\citenamefont {Cohen}\ and\ \citenamefont
  {Falicov}(1961)}]{Cohen1961}%
  \BibitemOpen
  \bibfield  {author} {\bibinfo {author} {\bibfnamefont {M.~H.}\ \bibnamefont
  {Cohen}}\ and\ \bibinfo {author} {\bibfnamefont {L.~M.}\ \bibnamefont
  {Falicov}},\ }\href {\doibase 10.1103/PhysRevLett.7.231} {\bibfield
  {journal} {\bibinfo  {journal} {Phys. Rev. Lett.}\ }\textbf {\bibinfo
  {volume} {7}},\ \bibinfo {pages} {231} (\bibinfo {year} {1961})}\BibitemShut
  {NoStop}%
\bibitem [{\citenamefont {Blount}(1962)}]{Blount1962}%
  \BibitemOpen
  \bibfield  {author} {\bibinfo {author} {\bibfnamefont {E.~I.}\ \bibnamefont
  {Blount}},\ }\href@noop {} {\bibfield  {journal} {\bibinfo  {journal} {Phys.
  Rev.}\ }\textbf {\bibinfo {volume} {126}},\ \bibinfo {pages} {1636} (\bibinfo
  {year} {1962})}\BibitemShut {NoStop}%
\bibitem [{\citenamefont {Shoenberg}(1984)}]{Shoenberg1984}%
  \BibitemOpen
  \bibfield  {author} {\bibinfo {author} {\bibfnamefont {D.}~\bibnamefont
  {Shoenberg}},\ }\href {\doibase 10.1017/CBO9780511897870} {\emph {\bibinfo
  {title} {{Magnetic Oscillations in Metals}}}},\ Cambridge Monographs on
  Physics\ (\bibinfo  {publisher} {Cambridge University Press},\ \bibinfo
  {year} {1984})\BibitemShut {NoStop}%
\bibitem [{\citenamefont {Kaganov}\ and\ \citenamefont
  {Slutskin}(1983)}]{Kaganov1983}%
  \BibitemOpen
  \bibfield  {author} {\bibinfo {author} {\bibfnamefont {M.~I.}\ \bibnamefont
  {Kaganov}}\ and\ \bibinfo {author} {\bibfnamefont {A.~A.}\ \bibnamefont
  {Slutskin}},\ }\href {\doibase 10.1016/0370-1573(83)90006-6} {\bibfield
  {journal} {\bibinfo  {journal} {Phys. Rep.}\ }\textbf {\bibinfo {volume}
  {98}},\ \bibinfo {pages} {189} (\bibinfo {year} {1983})}\BibitemShut
  {NoStop}%
\bibitem [{\citenamefont {van Delft}\ \emph {et~al.}(2018)\citenamefont {van
  Delft}, \citenamefont {Pezzini}, \citenamefont {Khouri}, \citenamefont
  {Mueller}, \citenamefont {Breitkreiz}, \citenamefont {Schoop}, \citenamefont
  {Carrington}, \citenamefont {Hussey},\ and\ \citenamefont
  {Wiedmann}}]{VanDelft2018}%
  \BibitemOpen
  \bibfield  {author} {\bibinfo {author} {\bibfnamefont {M.~R.}\ \bibnamefont
  {van Delft}}, \bibinfo {author} {\bibfnamefont {S.}~\bibnamefont {Pezzini}},
  \bibinfo {author} {\bibfnamefont {T.}~\bibnamefont {Khouri}}, \bibinfo
  {author} {\bibfnamefont {C.~S.~A.}\ \bibnamefont {Mueller}}, \bibinfo
  {author} {\bibfnamefont {M.}~\bibnamefont {Breitkreiz}}, \bibinfo {author}
  {\bibfnamefont {L.~M.}\ \bibnamefont {Schoop}}, \bibinfo {author}
  {\bibfnamefont {A.}~\bibnamefont {Carrington}}, \bibinfo {author}
  {\bibfnamefont {N.~E.}\ \bibnamefont {Hussey}}, \ and\ \bibinfo {author}
  {\bibfnamefont {S.}~\bibnamefont {Wiedmann}},\ }\href {\doibase
  10.1103/PhysRevLett.121.256602} {\bibfield  {journal} {\bibinfo  {journal}
  {Phys. Rev. Lett}\ }\textbf {\bibinfo {volume} {121}},\ \bibinfo {pages}
  {256602} (\bibinfo {year} {2018})}\BibitemShut {NoStop}%
\bibitem [{\citenamefont {Sebastian}\ \emph {et~al.}(2012)\citenamefont
  {Sebastian}, \citenamefont {Harrison}, \citenamefont {Liang}, \citenamefont
  {Bonn}, \citenamefont {Hardy}, \citenamefont {Mielke},\ and\ \citenamefont
  {Lonzarich}}]{Sebastian2012}%
  \BibitemOpen
  \bibfield  {author} {\bibinfo {author} {\bibfnamefont {S.~E.}\ \bibnamefont
  {Sebastian}}, \bibinfo {author} {\bibfnamefont {N.}~\bibnamefont {Harrison}},
  \bibinfo {author} {\bibfnamefont {R.}~\bibnamefont {Liang}}, \bibinfo
  {author} {\bibfnamefont {D.~A.}\ \bibnamefont {Bonn}}, \bibinfo {author}
  {\bibfnamefont {W.~N.}\ \bibnamefont {Hardy}}, \bibinfo {author}
  {\bibfnamefont {C.~H.}\ \bibnamefont {Mielke}}, \ and\ \bibinfo {author}
  {\bibfnamefont {G.~G.}\ \bibnamefont {Lonzarich}},\ }\href {\doibase
  10.1103/PhysRevLett.108.196403} {\bibfield  {journal} {\bibinfo  {journal}
  {Phys. Rev. Lett.}\ }\textbf {\bibinfo {volume} {108}},\ \bibinfo {pages}
  {196403} (\bibinfo {year} {2012})}\BibitemShut {NoStop}%
\bibitem [{\citenamefont {Dwivedi}(2018)}]{Dwivedi2018}%
  \BibitemOpen
  \bibfield  {author} {\bibinfo {author} {\bibfnamefont {V.}~\bibnamefont
  {Dwivedi}},\ }\href {\doibase 10.1103/PhysRevB.97.064201} {\bibfield
  {journal} {\bibinfo  {journal} {Phys. Rev. B}\ }\textbf {\bibinfo {volume}
  {97}},\ \bibinfo {pages} {064201} (\bibinfo {year} {2018})}\BibitemShut
  {NoStop}%
\bibitem [{\citenamefont {Abdulla}\ \emph {et~al.}(2021)\citenamefont
  {Abdulla}, \citenamefont {Rao},\ and\ \citenamefont {Murthy}}]{Abdulla2021}%
  \BibitemOpen
  \bibfield  {author} {\bibinfo {author} {\bibfnamefont {F.}~\bibnamefont
  {Abdulla}}, \bibinfo {author} {\bibfnamefont {S.}~\bibnamefont {Rao}}, \ and\
  \bibinfo {author} {\bibfnamefont {G.}~\bibnamefont {Murthy}},\ }\href
  {\doibase 10.1103/PhysRevB.103.235308} {\bibfield  {journal} {\bibinfo
  {journal} {Phys. Rev. B}\ }\textbf {\bibinfo {volume} {103}},\ \bibinfo
  {pages} {235308} (\bibinfo {year} {2021})}\BibitemShut {NoStop}%
\bibitem [{\citenamefont {Mathur}\ \emph {et~al.}()\citenamefont {Mathur},
  \citenamefont {Yuan}, \citenamefont {Cheng}, \citenamefont {Kaushik},
  \citenamefont {Robredo},\ and\ \citenamefont {Maia}}]{Mathur2022}%
  \BibitemOpen
  \bibfield  {author} {\bibinfo {author} {\bibfnamefont {N.}~\bibnamefont
  {Mathur}}, \bibinfo {author} {\bibfnamefont {F.}~\bibnamefont {Yuan}},
  \bibinfo {author} {\bibfnamefont {G.}~\bibnamefont {Cheng}}, \bibinfo
  {author} {\bibfnamefont {S.}~\bibnamefont {Kaushik}}, \bibinfo {author}
  {\bibfnamefont {I.}~\bibnamefont {Robredo}}, \ and\ \bibinfo {author}
  {\bibfnamefont {G.}~\bibnamefont {Maia}},\ }\href@noop {} {\ }\Eprint
  {http://arxiv.org/abs/2212.13688} {arXiv:2212.13688} \BibitemShut {NoStop}%
\bibitem [{\citenamefont {Kaushik}\ \emph {et~al.}()\citenamefont {Kaushik},
  \citenamefont {Robredo}, \citenamefont {Mathur}, \citenamefont {Schoop},
  \citenamefont {Jin}, \citenamefont {Vergniory},\ and\ \citenamefont
  {Cano}}]{Kaushik2022}%
  \BibitemOpen
  \bibfield  {author} {\bibinfo {author} {\bibfnamefont {S.}~\bibnamefont
  {Kaushik}}, \bibinfo {author} {\bibfnamefont {I.}~\bibnamefont {Robredo}},
  \bibinfo {author} {\bibfnamefont {N.}~\bibnamefont {Mathur}}, \bibinfo
  {author} {\bibfnamefont {L.~M.}\ \bibnamefont {Schoop}}, \bibinfo {author}
  {\bibfnamefont {S.}~\bibnamefont {Jin}}, \bibinfo {author} {\bibfnamefont
  {M.~G.}\ \bibnamefont {Vergniory}}, \ and\ \bibinfo {author} {\bibfnamefont
  {J.}~\bibnamefont {Cano}},\ }\href {http://arxiv.org/abs/2207.14109} {\
  }\Eprint {http://arxiv.org/abs/2207.14109} {arXiv:2207.14109} \BibitemShut
  {NoStop}%
\bibitem [{\citenamefont {Landau}\ and\ \citenamefont
  {Lifshitz}(1977)}]{Landau1977}%
  \BibitemOpen
  \bibfield  {author} {\bibinfo {author} {\bibfnamefont {L.~D.}\ \bibnamefont
  {Landau}}\ and\ \bibinfo {author} {\bibfnamefont {E.~M.}\ \bibnamefont
  {Lifshitz}},\ }\href@noop {} {\emph {\bibinfo {title} {{Course of Theoretical
  Physics}}}}\ (\bibinfo  {publisher} {Elsevier},\ \bibinfo {address}
  {Oxford},\ \bibinfo {year} {1977})\BibitemShut {NoStop}%
\bibitem [{\citenamefont {Ramshaw}\ \emph {et~al.}(2018)\citenamefont
  {Ramshaw}, \citenamefont {Modic}, \citenamefont {Shekhter}, \citenamefont
  {Zhang}, \citenamefont {Kim}, \citenamefont {Moll}, \citenamefont {Bachmann},
  \citenamefont {Chan}, \citenamefont {Betts}, \citenamefont {Balakirev},
  \citenamefont {Migliori}, \citenamefont {Ghimire}, \citenamefont {Bauer},
  \citenamefont {Ronning},\ and\ \citenamefont {McDonald}}]{Ramshaw2018}%
  \BibitemOpen
  \bibfield  {author} {\bibinfo {author} {\bibfnamefont {B.~J.}\ \bibnamefont
  {Ramshaw}}, \bibinfo {author} {\bibfnamefont {K.~A.}\ \bibnamefont {Modic}},
  \bibinfo {author} {\bibfnamefont {A.}~\bibnamefont {Shekhter}}, \bibinfo
  {author} {\bibfnamefont {Y.}~\bibnamefont {Zhang}}, \bibinfo {author}
  {\bibfnamefont {E.-A.}\ \bibnamefont {Kim}}, \bibinfo {author} {\bibfnamefont
  {P.~J.~W.}\ \bibnamefont {Moll}}, \bibinfo {author} {\bibfnamefont {M.~D.}\
  \bibnamefont {Bachmann}}, \bibinfo {author} {\bibfnamefont {M.~K.}\
  \bibnamefont {Chan}}, \bibinfo {author} {\bibfnamefont {J.~B.}\ \bibnamefont
  {Betts}}, \bibinfo {author} {\bibfnamefont {F.}~\bibnamefont {Balakirev}},
  \bibinfo {author} {\bibfnamefont {A.}~\bibnamefont {Migliori}}, \bibinfo
  {author} {\bibfnamefont {N.~J.}\ \bibnamefont {Ghimire}}, \bibinfo {author}
  {\bibfnamefont {E.~D.}\ \bibnamefont {Bauer}}, \bibinfo {author}
  {\bibfnamefont {F.}~\bibnamefont {Ronning}}, \ and\ \bibinfo {author}
  {\bibfnamefont {R.~D.}\ \bibnamefont {McDonald}},\ }\href {\doibase
  10.1038/s41467-018-04542-9} {\bibfield  {journal} {\bibinfo  {journal} {Nat.
  Commun.}\ }\textbf {\bibinfo {volume} {9}},\ \bibinfo {pages} {2217}
  (\bibinfo {year} {2018})}\BibitemShut {NoStop}%
\bibitem [{\citenamefont {Dwivedi}\ and\ \citenamefont
  {Chua}(2016)}]{Dwivedi2016b}%
  \BibitemOpen
  \bibfield  {author} {\bibinfo {author} {\bibfnamefont {V.}~\bibnamefont
  {Dwivedi}}\ and\ \bibinfo {author} {\bibfnamefont {V.}~\bibnamefont {Chua}},\
  }\href {\doibase 10.1103/PhysRevB.93.134304} {\bibfield  {journal} {\bibinfo
  {journal} {Phys. Rev. B}\ }\textbf {\bibinfo {volume} {93}},\ \bibinfo
  {pages} {134304} (\bibinfo {year} {2016})}\BibitemShut {NoStop}%
\bibitem [{\citenamefont {Groth}\ \emph {et~al.}(2014)\citenamefont {Groth},
  \citenamefont {Wimmer}, \citenamefont {Akhmerov},\ and\ \citenamefont
  {Waintal}}]{Groth2014}%
  \BibitemOpen
  \bibfield  {author} {\bibinfo {author} {\bibfnamefont {C.~W.}\ \bibnamefont
  {Groth}}, \bibinfo {author} {\bibfnamefont {M.}~\bibnamefont {Wimmer}},
  \bibinfo {author} {\bibfnamefont {A.~R.}\ \bibnamefont {Akhmerov}}, \ and\
  \bibinfo {author} {\bibfnamefont {X.}~\bibnamefont {Waintal}},\ }\href
  {\doibase 10.1088/1367-2630/16/6/063065} {\bibfield  {journal} {\bibinfo
  {journal} {New J. Phys.}\ }\textbf {\bibinfo {volume} {16}},\ \bibinfo
  {pages} {063065} (\bibinfo {year} {2014})}\BibitemShut {NoStop}%
\bibitem [{\citenamefont {Dwivedi}\ and\ \citenamefont
  {Ramamurthy}(2016)}]{Dwivedi2016}%
  \BibitemOpen
  \bibfield  {author} {\bibinfo {author} {\bibfnamefont {V.}~\bibnamefont
  {Dwivedi}}\ and\ \bibinfo {author} {\bibfnamefont {S.~T.}\ \bibnamefont
  {Ramamurthy}},\ }\href {\doibase 10.1103/PhysRevB.94.245143} {\bibfield
  {journal} {\bibinfo  {journal} {Phys. Rev. B}\ }\textbf {\bibinfo {volume}
  {94}},\ \bibinfo {pages} {245143} (\bibinfo {year} {2016})}\BibitemShut
  {NoStop}%
\bibitem [{\citenamefont {Breitkreiz}\ and\ \citenamefont
  {Brouwer}()}]{Breitkreiz2022}%
  \BibitemOpen
  \bibfield  {author} {\bibinfo {author} {\bibfnamefont {M.}~\bibnamefont
  {Breitkreiz}}\ and\ \bibinfo {author} {\bibfnamefont {P.~W.}\ \bibnamefont
  {Brouwer}},\ }\href {\doibase 10.48550/ARXIV.2212.11059} {\enquote {\bibinfo
  {title} {{Fermi-arc metals}},}\ }\Eprint {http://arxiv.org/abs/2212.11059}
  {arXiv:2212.11059} \BibitemShut {NoStop}%
\end{thebibliography}%


\onecolumngrid
\section*{SUPPLEMENTAL MATERIAL}
\twocolumngrid
 
\renewcommand{\theequation}{S\arabic{equation}}
\renewcommand{\thefigure}{S\arabic{figure}}
\setcounter{equation}{0}
\setcounter{figure}{0}

\section*{Topological argument for homochiral interface Fermi arcs} 

\begin{figure}
    \includegraphics[width= 0.67\columnwidth]{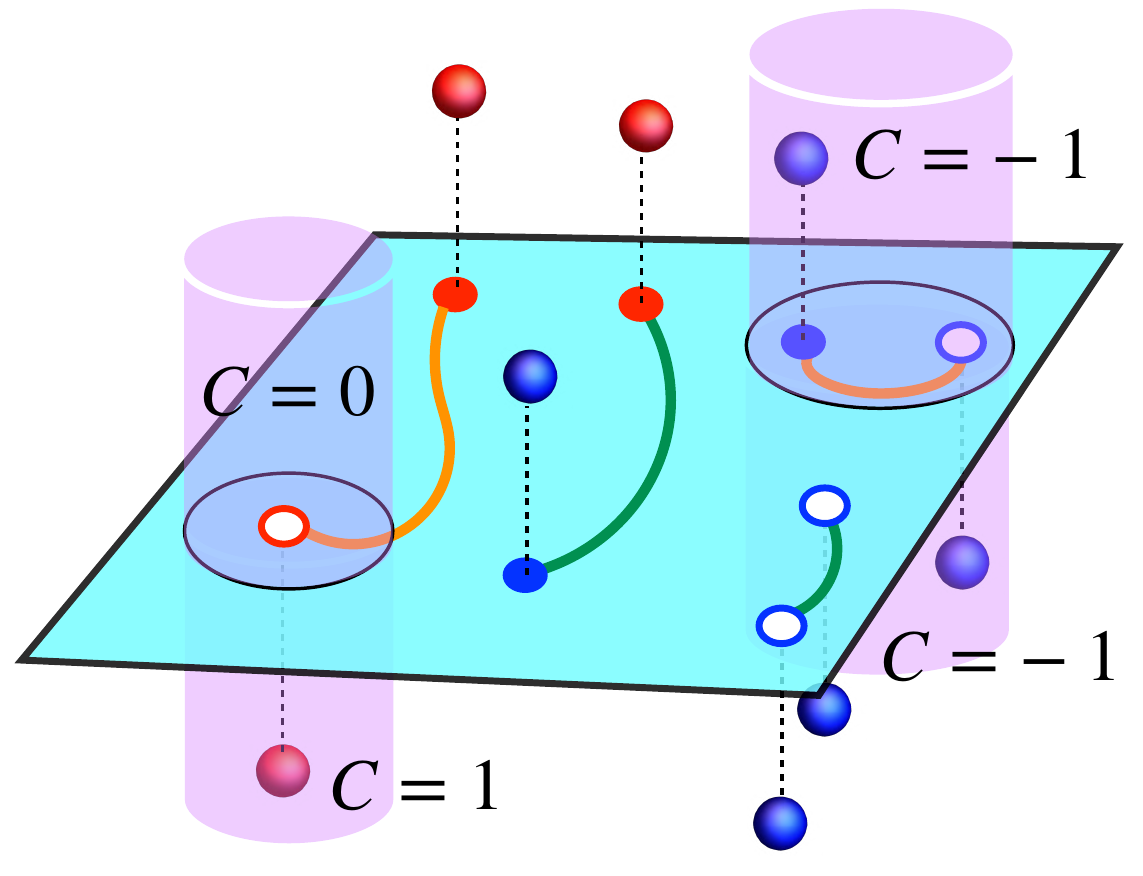}  $\;$ 
    \includegraphics[width=0.28\columnwidth]{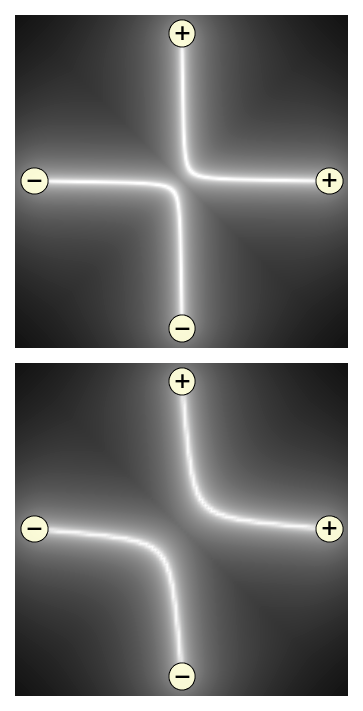} 
    \caption{
      (Left) Interface between two WSMs with Weyl nodes of positive/negative chirality (red/blue spheres) and their projections onto the interface from the upper WSM (filled circles) and lower WSM (empty circles). Fermi arcs can either connect Weyl nodes of opposite chirality on the same side of the interface (green lines) or same chirality on opposite sides of the interface (orange lines), corresponding to heterochiral and homochiral connectivities, respectively. The topology of interface Fermi arcs can be inferred from considering the interface of 2D Chern insulators with Chern number $C$, indicated as cylinders.  
      (Right) Numerically computed interface Fermi arcs for the lattice model Eq.~\eqref{etas} with $\tht = \pi/2$ for $\kappa = 0.1$ (top) and $\kappa=0.25$ (bottom). }
      \label{int_loops}
\end{figure}

Consider an interface between two WSMs whose Weyl nodes occur at different transverse lattice momenta, as shown in Fig.\ \ref{int_loops}. We begin by noting that the the restriction of the lattice model to transverse momenta forming a loop in the interface away from the projections of the Weyl nodes can be interpreted as the band structure of two semi-infinite gapped 2d systems, which can possess a Chern number. An interface between two gapped phases with Chern numbers $C_1$ and $C_2$ hosts $\abs{C_1-C_2}$ protected chiral modes in the bulk gap. Thus, for a given loop, we can infer the number of topologically protected zero modes crossing it, depending on the enclosed Weyl nodes on either side of the interface. In particular, if the loop encloses a single Weyl node of chirality $\chi$ from one of the WSMs and none from the other, then the loop corresponds to an interface between gapped 2D phases with Chern numbers $\chi$ and $0$, guaranteeing the existence of a chiral mode crossing zero energy at some momentum on the loop. On the other hand, if the loop encloses Weyl nodes of identical chirality $\chi$ from both WSMs, it corresponds to an interface between two gapped phases with identical Chern numbers, which does not host any topologically protected modes. 

In terms of the interface between Fermi arcs, we thus conclude that a Fermi arc emanate from each projection of a Weyl node, but given a loop enclosing Weyl nodes of identical chiralities from the two sides, the Fermi arcs can always be deformed to lie entirely inside it. Thus, the topologically protected interface Fermi arcs must connect projections of Weyl nodes of identical chiralities coming from the two WSMs. 

\section*{Tunnel conductance from Landauer approach}

\begin{figure}
\includegraphics[width=0.7\columnwidth]{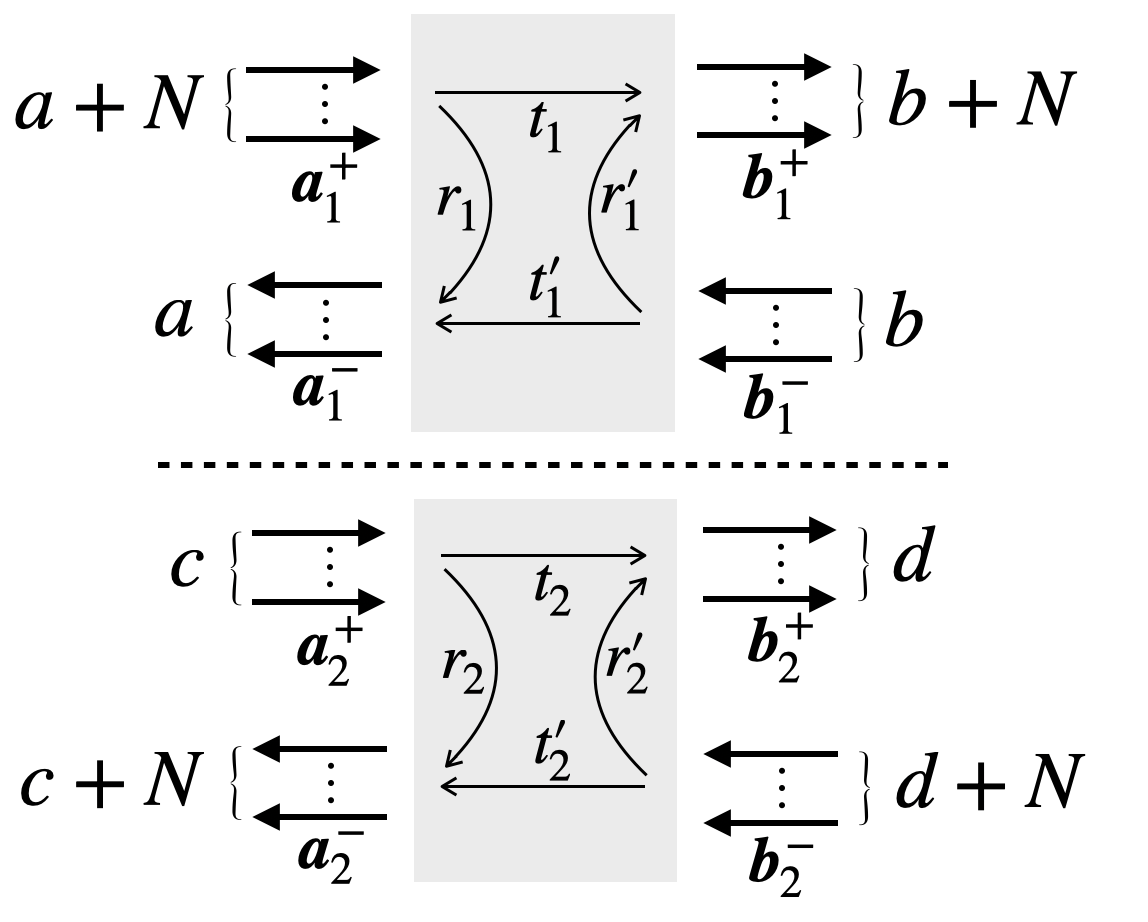}
\caption{Illustration of the scattering problem, Eq.\ \eqref{s}, indicating the numbers of modes.}
\label{figsm}
\end{figure}

The Landauer formula $G=(e^2/h)T$ relates the conductance $G$ with the transmission probability $T$, the sum of transmission probabilities over all right- or leftmoving modes. We consider the transmission of a flat interface where the modes (scattering states) are divided into two decoupled groups, labeled $1$ and $2$, whereby in group $1$ the number of rightmovers is larger by $N$ modes than the number of leftmovers, and vice versa for group $2$. The scattering problem, illustrated in Fig.\ \ref{figsm}, with scattering matrices $S_i$ for the two groups $i=1,2$, reads
\begin{equation}
\begin{pmatrix} \bm{a}^-_i \\  \bm{b}^+_i \end{pmatrix} = \underbrace{
\begin{pmatrix} r_i & t_i' \\ t_i & r_i' \end{pmatrix}}_{\equiv S_i}
\begin{pmatrix}
\bm{a}^+_i \\ \bm{b}^-_i\end{pmatrix}.
\label{s} 
\end{equation}
Note that the reflection and transmission 
amplitude matrices have different rectangular shapes, set by the size of the vectors of mode coefficients.

We will now show that the total transmission probability is bound from below as
\begin{equation}
    T \equiv \sum_i\mathrm{Tr}\big(t^\dagger_i t^\pd_i \big) \geq N.
    \label{trp}
\end{equation}
Heuristically one can understand from particle conservation that, considering e.g., the modes to the left of the interface, only $a$ of the $a+N$ rightmovers in subsystem $1$ can be reflected since there are only $a$ leftmovers and, in subsystem 2, all $c$ rightmovers can be reflected since there are $c+N$ leftmovers. Hence, in total, at least $N$ modes must be transmitted. (The same conclusion can be reached considering the modes to the right of the interface.) Formally, this derives from the
unitarity of the scattering matrices, $t^\dagger_1 t^\pd_1+r^\dagger_1 r^\pd_1=\mathbb{1}_{(a+N)\times(a+N)} $ and 
$t'_1 {t^{\prime}_1}^\dagger +r_1 r^{\dagger}_1=\mathbb{1}_{a\times a} $. Taking the 
trace of both equations and eliminating 
$\mathrm{Tr}\big(r^\dagger_1 r^\pd_1)$ and repeating the same for the other subsystem, we obtain 
\begin{align}
\mathrm{Tr}\big(t^\dagger_1 t^\pd_1)=&\  
\mathrm{Tr}\big({t^{\prime}_1}^\dagger {t^{\prime}_1}^\pd\big)+N\\
\mathrm{Tr}\big(t^\dagger_2 t^\pd_2)=&\  
\mathrm{Tr}\big({t^{\prime}_2}^\dagger {t^{\prime}_2}^\pd\big)-N.
\end{align}
Equation \eqref{trp} follows by taking into account 
 $\mathrm{Tr}\big({t^{(\prime)\dagger}_i} {t^{(\prime)}_i}^\pd\big)\geq 0$.

We can apply this result to a pair of chiral 
Weyl Fermions in an applied magnetic field directed perpendicular
to the interface, at which the homochiral Fermi arcs connect 
Weyl Fermions of the same chirality at opposite sides of the interface. In this case, the difference in the number of right- and 
leftmovers is the degeneracy of the chiral Landau level $N=N(B)$ and subsystem $1$ and $2$ correspond 
to positive and negative chirality.
The lower bound of the conductance for a single 
homochiral Fermi arc is thus
$ G = \frac{e^2}{h}T \geq  \frac{e^2}{h} N(B)$.
The upper bound is set by the number of states at the smallest 
constriction. As argued in the main text, all particles must pass the Fermi arc, which in the magnetic field becomes a single $N(B)$-degenerate Landau level. It follows that
\begin{equation}
    G =  n_\mathrm{ho} N(B) \frac{e^2}{h},
\end{equation}
where we included the total number $n_\mathrm{ho}$ of homochiral
connectivities.

\section*{Tunnel conductance from semiclassics}

In an applied magnetic field directed perpendicular
to the interface, the Lorentz force moves a particle along the Fermi arc in time $dt$  by the amount $dk = vdt/\ell_B^2$, where $v$ is the velocity and $\ell_B$ the magnetic length. Particles contributing to 
transport are within the energy range $eU$, set by the applied 
voltage $U$, which corresponds to the momentum $eU/\hbar v$. 
Hence, the transported charge is
\begin{equation}
dQ = e\frac{dk\,(eU/\hbar v)}{(2\pi)^2}{\cal A},
\end{equation}
where ${\cal A}$ is the interface area. Inserting $dk = vdt/\ell_B^2$
and with $N(B)= {\cal A}/(2\pi \ell_B^2)$ we obtain
\begin{equation}
  \frac{dQ}{Udt} = N(B) \frac{e^2}{h} .
\end{equation}
This correctly corresponds to the tunneling conductance across the 
interface via a single homochiral Fermi arc obtained from the scattering-matrix formalism.

\section*{Interface modes of a lattice model using transfer matrices}

The generalized transfer matrix along $x$ for the two-band models lattice model of Eq.~\eqref{eq:hlt_orig} is given by \cite{Dwivedi2016, Dwivedi2016b}
\begin{equation} 
  T(\ve, \vk_\perp) = 
  \frac1{1+\eta_z}
  \begin{pmatrix}
    \ve^2 - \Gamma^2  &   -\ve+\eta_y \\ 
    \ve+\eta_y        & -1 
  \end{pmatrix}, 
  \label{TM}
\end{equation} 
where $\Gamma \equiv \eta_y^2 + (1 + \eta_z)^2$. 
For an interface along $x$ with tunneling strength $\kappa$, a necessary condition for the existence of a mode localized at the interface is \cite{Dwivedi2018}
\begin{equation} 
  \det \left[ T_+, K T_- K \right] = 0, 
\end{equation} 
where $T_\pm $ denote the transfer matrices on the two sides of the interface and $K = \diag{(1/\kappa) ,\kappa }$. For the transfer matrix Eq.\ \eqref{TM}, this expression becomes
\begin{equation} 
  \Lambda_+ - \Lambda_- = \pm 2\sinh\lambda \sqrt{ \Lambda_+ \Lambda_- + 4 \cosh^2\lambda }, 
  \label{int_arcs}
\end{equation}  
where $\lambda = \ln \kappa$ and 
\begin{equation} 
  \Lambda_\pm 
  = \frac1{\eta_y^\pm} \left[ 1 - \left( \eta_y^\pm \right)^2 - \left( 1 + \eta_z^\pm \right)^2 \right].
  \label{int_arcs2}
\end{equation}
For $\Lambda_+ = -\Lambda_- \equiv \Lambda$, Eq.~\eqref{int_arcs}  simplifies to 
\begin{equation} 
  \Lambda = \pm2\sinh\lambda = \pm \left( \kappa - \frac1\kappa\right). 
  \label{int_arcs_simplified}
\end{equation}  
This turns out to be the relevant limit for our setup. 

For the interface between WSMs described by Eq.\ \eqref{etas}, linearizing around $\vk_\perp = \nullv$ yields
\begin{align}
  \eta^{\pm}_y(\vk_\perp) &= \pm v_y k_y + v_z k_z, \nonumber \\ 
  \eta^{\pm}_z(\vk_\perp) &= \beta + \frac12 \left( k_y^2 + k_z^2 \right),
\end{align}
where $v_i=\sin b_i/\sin b$ and $\beta=\cos(b_y) + \cos(b_z) - 2$. Thus, the Fermi arcs at the interface for $\kappa = 0$ lie along $v_z k_z = \pm v_y k_y$, leading to a crossing at $\vk_\perp = 0$. For $\kappa>0$, the interface Fermi arc can be computed using Eqns.~\eqref{int_arcs} and \eqref{int_arcs2}. In particular, the minimum separation between the arcs lies along $k_y$. Setting $k_z = 0$, we get 
\begin{equation}
    \Lambda_\pm = \pm \frac1{v_y k_y} \left[1 - (v_y k_y)^2 - \left(1 + \beta + \frac{k_y^2}2  \right)^2 \right]
\end{equation}
so that $\Lambda_+ = -\Lambda_-$. The zero crossings along $k_z=0$ and thus the minimum separation between the interface Fermi arcs are given by Eq.~\ref{int_arcs_simplified}. For $\kappa \ll 1$, we only retain the terms at $\O(1/\kappa)$ and $\O(1/k_y)$, so that 
\begin{equation}
    \frac1\kappa 
    = -\frac1{v_y k_y} \left[1 - \left(1 + \beta \right)^2 \right]
    = \frac{\beta(2+\beta)}{v_y k_y}.
\end{equation}
The minimum separation between the interface Fermi arc at $\O(\kappa)$ is thus given by 
\begin{equation} 
  \Delta  = 2 k_y = \frac{2\kappa \beta (2+\beta)}{v_y}. 
\end{equation}

\end{document}